\documentstyle[11pt,paspconf]{article}
\input epsf.sty
\begin{document}

\title{The artificial sky brightness in Europe derived from DMSP satellite data.}

\author{P. Cinzano and F. Falchi}
\affil{Dipartimento di Astronomia, Universit\`a di Padova,
             vicolo dell'Osservatorio 5,  I-35122 Padova, Italy}

\author{C. D. Elvidge and K. E. Baugh}
\affil{Solar-Terrestrial Physics Division, NOAA National Geophysical Data Center, 3100 Marine Street, Boulder CO 80303}


\altaffiltext{1}{email: cinzano@pd.astro.it}



\begin{abstract}
We present the map of the artificial sky brightness in Europe in V band with a resolution of approximately 1 km. The aim is to understand the state of night sky pollution in Europe, to quantify the present situation and to allow future monitoring of trends.

The artificial sky brightness in each site at a given position on the sky is obtained by integration of the contributions produced by every surface area in the surroundings of the site. Each contribution is computed taking into account based on detailed models the propagation in the atmosphere of the upward light flux emitted by the area and measured by the Operational Linescan System of DMSP satellites.  The modelling technique, introduced and developed by Garstang  and also applied by Cinzano, takes into account the extinction along light paths, a double scattering of light from atmospheric molecules and aerosols, Earth curvature and  allows to associate the predictions to the aerosol content of the atmosphere. 

\end{abstract}


\keywords{atmospheric effects,
               site testing,
               scattering, 
                artificial sky brightness, light pollution}


\section{Introduction}

An effective battle against light pollution requires the knowledge of the situation of the night sky in large territories, the recognition of the most concerned areas, the determination of the growth trends, the identification of more polluting cities. Therefore  a method to map the artificial sky brightness in large territories is required. This is also useful in order to recognize less polluted areas and potential astronomical sites.

DMSP satellite allows direct information on the upward light emission from almost all countries around the World (Sullivan 1989; Elvidge et al. 1997a, 1997b, 1997c, 1999; Isobe \& Hamamura 1998).
We present the outlines of a method to  map the artificial sky brightness in large territories   measuring the upward flux in DMSP satellite night-time images in order to bypass errors  arising when using population data to estimate upward flux, and computing its effects with detailed modelling of light pollution propagation in the atmosphere. Details will be extensively discussed in a forecoming paper (Cinzano et al. 1999, in prep.).

\section{Satellite data}
\label{s1}

U.S. Air Force Defense Meteorological Satellite Program (DMSP) satellites are in low altitude (830 km) sun/synchronous polar orbits with an orbital period of 101 minutes. With 14 orbits per day they generate a global nightime and daytime coverage of the Earth every 24 hours. 
The Operational Linescan System (OLS) is an oscillating scan radiometer with low-light visible and thermal infrared imaging capabilities. 
At night the instrument for visible imagery is a Photo Multiplier Tube (PMT) sensitive to radiation from 470 nm to 900 nm  FWHM with the highest sensitivity at 550-650 nm where the most widely used lamps for external night-time lighting have the strongest emission.
Most of data received by National Oceanic and Atmospheric Administration (NOAA) National Geophysics Data Center (NGDC), which archives DMSP data since 1992, are smoothed by on-board averaging of 5 by 5 adjacent detector pixels and have a nominal space resolution of 2.8 km. 

In three observational runs made during the darkest portions of lunar cycles during March of 1996 plus January and February of 1997, NGDC acquired OLS data at reduced gain settings in order to avoid saturation produced in a large number of pixels inside cities in normal gain operations due at high OLS-PMT sensivity. 
Three different gain settings were used on alternating nights to overcome the dynamic range limitations of the OLS.
With these data a cloud-free radiance calibrated composite image of the Earth (Elvidge et al. 1999) has been obtained. 
The temporal compositing makes it possible to remove noise and lights from ephemeral events such as fire and lightning.
Main steps in the nighttime lights product generation are:
1) establishment of a reference grid with finer spatial resolution than the input imagery;
2) identification of the cloud free section of each orbit based on OLS thermal band data;
3) identification of lights and removal of noise and solar glare;
4) projection of the lights from cloud-free areas from each orbit into the reference grid, with calibration to radiance units;
5) tallying of the total number of light detections in each grid cell and calculation of the average radiance value;
6) filtering images based on frequency of detection to remove ephemeral events.
The final image was transformed in a latitude/longitude projection with 30''x30'' pixel size. The map of Europe was obtained with a portion of 5000x5000 pixel of this final image, starting at longitude 10$\deg$ 30' west and   latitude 72\deg north. 

\section{Mapping technique}
\label{s4}

Scattering from atmospheric particles and molecules spreads the light emitted upward by the sources. If $e(x,\,y)$ is the upward emission per unit area in $(x,\,y)$, the total artificial sky brightness in a given direction of the sky in a site in  $(x',\,y')$ is:
\begin{equation}
\label{int1}
b(x',\,y')=\int\int e(x,y) f((x,y),(x',\,y'))~dx ~dy
\end{equation}
where $f((x,\,y),(x',\,y'))$ give the artificial sky brightness per unit of upward light emission produced by the unitary area in $(x,\,y)$ in the site in $(x',\,y')$. The light pollution propagation function $f$ depends in general on the geometrical disposition (altitude of the site and the area, and their mutual distance), on the atmospheric distribution of molecules and aerosols and their optical characteristics in the choosen photometrical band, on the shape of the emission function of the source and on the direction of the sky observed. In some works this function has been approximated with a variety of semi-empirical propagation law like Treanor Law (Treanor 1973; Falchi and Cinzano 1999; Cinzano and Falchi 1999), Walker Law (Walker 1973), Berry Law (Berry 1976), Garstang Law (Garstang 1991b). However, all of them do not take into account the effects of Earth curvature that cannot be neglected in accurate mapping of large and non-isolated territories.

We obtained the propagation function   $f((x,y),(x',\,y'))$ for each couple of points $(x,y)$ and $(x',\,y')$   with detailed models for the light propagation in the atmosphere based on the modelling technique introduced and developed by Garstang  (1986, 1987, 1988, 1989a, 1989b, 1991a, 1991b, 1991c, 1999)  and also applied by Cinzano (1999a, 1999b, 1999c). The models assume Rayleigh scattering by molecules and Mie scattering by aerosols and take into account extinction along light path and Earth curvature. They allow to associate the predictions to well-defined parameters related to the aerosol content, so the atmospheric conditions at which predictions refer can be well known. 
Depending results on an integration over a large zone, the resolution of the maps is better than resolution of the original images and is generally of the order of the distance between two pixel centers (less than 1km). However where sky brightness is dominated by contribution of nearest land areas, effects of the resolution of the original image could became relevant.

We assumed the atmosphere in hydrostatic equilibrium under the gravitational force and an exponential decrease of number density for the atmospheric haze aerosols. Measurements show  that for the first 10 km this is a reasonable approximation.
We are interested at average atmospheric conditions, better if typical and not at particular conditions of a given night, so a detailed modelling of local aerosol distribution at a given night is beyond the scope of this work.  We neglected presence of sporadic denser aerosol layers at various heights or at ground level, the effects of the Ozone layer and the presence of volcanic dust. 
We take into account changes in aerosol content as Garstang (1986) introducing a parameter  $K$ which measures the relative importance of aerosol and molecules for scattering light. The adopted modelling technique allows to assess the atmospheric conditions for which a map is computed giving observable quantities like the vertical extinction at sea level  in magnitudes.
More detailed atmospheric models could be used whenever available.

The angular scattering function for atmospheric haze aerosols can be measured easily with a number of well known remote-sensing techniques.  Being interested in a typical average function, we adopted the same function used by Garstang (1991a) and we neglected geographical gradients.
The normalized emission function of each area  gives the relative upward flux per unit solid angle in each direction. It is the sum of the direct emission from fixtures and the reflected emission from lighted  surfaces, normalized to its integral and is not known. 
In this paper we assumed that all land areas have the same average normalized emission function. This is equivalent to assuming that lighting habits are similar on average in each land area and  that differences from the average are casually distributed in the territory.
We choose to assume this function and check its consistency with satellite measurements, rather than directly determine it from satellite measurements because at very low elevation angles the spread is too much large to constrain adequately the function shape.
We adopted for the average normalized emission function the 
normalized city emission function from Garstang (1986).

\section{Results}
\label{s6}

\begin{figure}
\epsfysize=7.8cm 
\hspace{2.0cm}\epsfbox{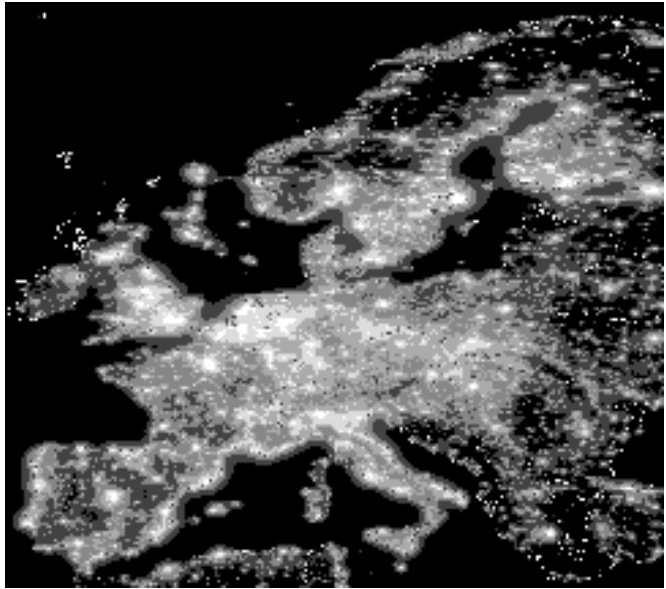} 
\caption[h]{Artificial sky brightness at sea level in Europe in V band  for aerosol content parameter $K\!=\!1$.}
\label{res1}
\end{figure}

\begin{figure}
\epsfysize=10.4cm 
\hspace{0.8cm}\epsfbox{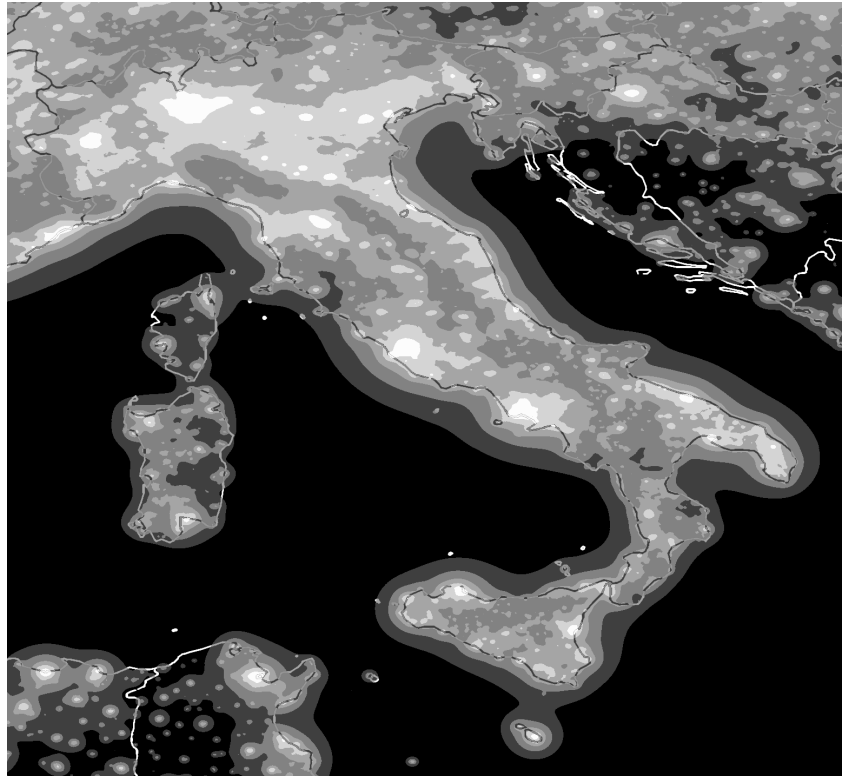} 
\caption[h]{Artificial sky brightness at sea level in Italy in V band  for aerosol content parameter $K\!=\!1$. }
\label{res2}
\end{figure}

\begin{figure}
\epsfysize=9.7cm 
\hspace{0.8cm}\epsfbox{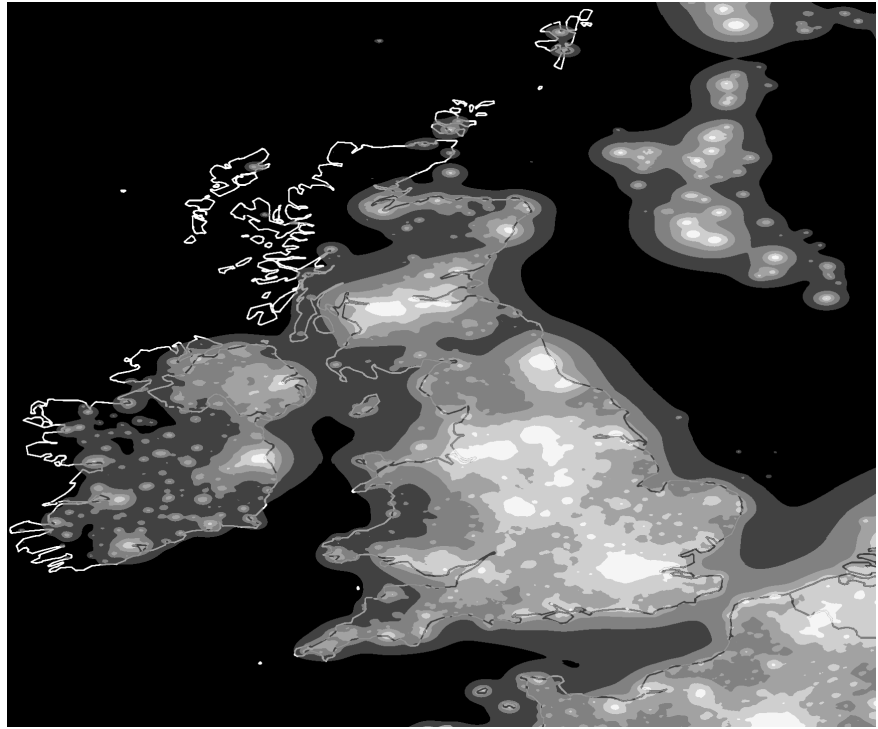} 
\caption[h]{Artificial sky brightness at sea level in Great Britain and Ireland in V band  for aerosol content parameter $K\!=\!1$. }
\label{res3}
\end{figure}

\begin{figure}
\epsfysize=8.5cm 
\hspace{0.8cm}\epsfbox{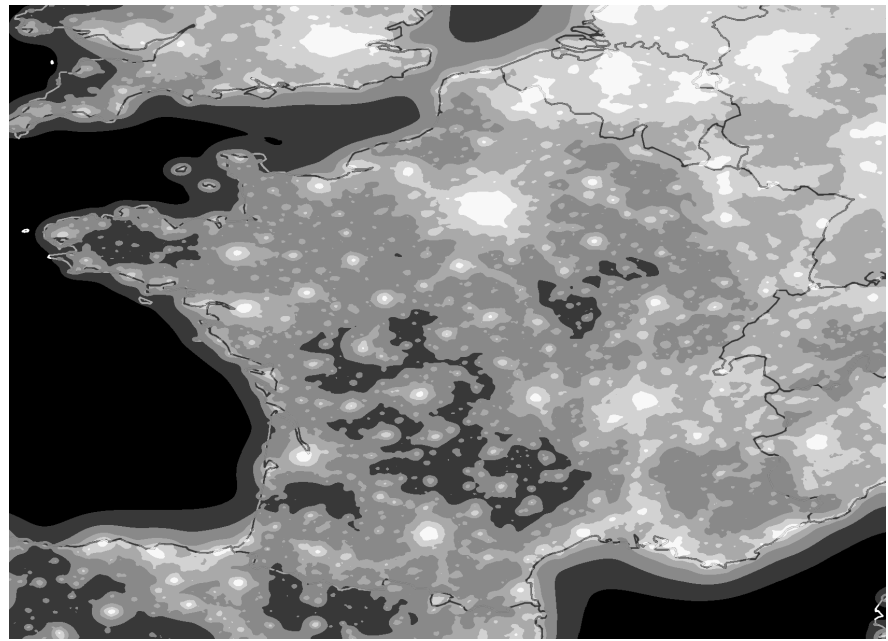} 
\caption[h]{Artificial sky brightness at sea level in France and Belgium in V band  for aerosol content parameter $K\!=\!1$. }
\label{res4}
\end{figure}

\begin{figure}
\epsfysize=8.7cm 
\hspace{0.7cm}\epsfbox{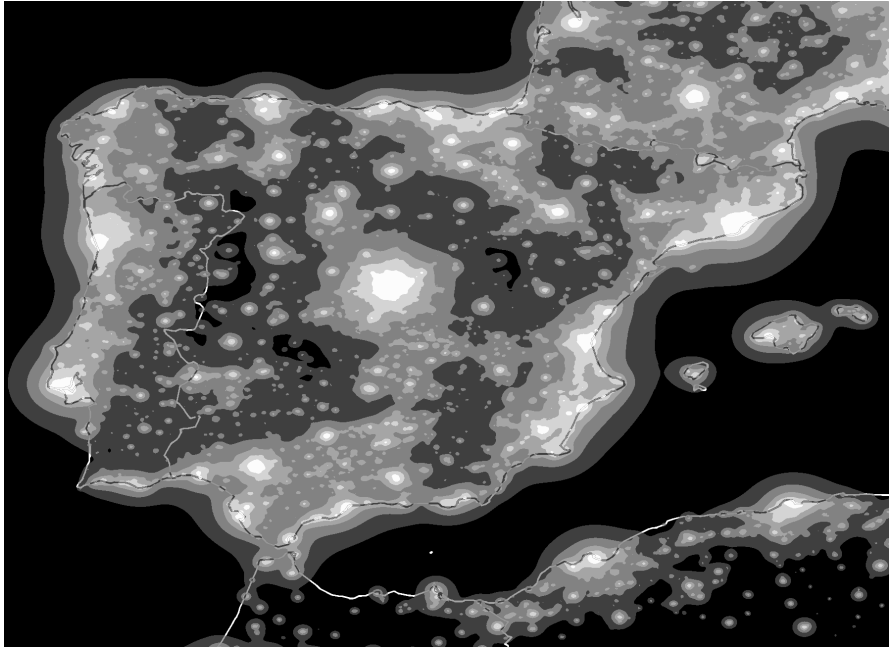} 
\caption[h]{Artificial sky brightness at sea level in Spain and Portugal in V band  for aerosol content parameter $K\!=\!1$. }
\label{res5}
\end{figure}

\begin{figure}
\epsfysize=9.5cm 
\hspace{0.7cm}\epsfbox{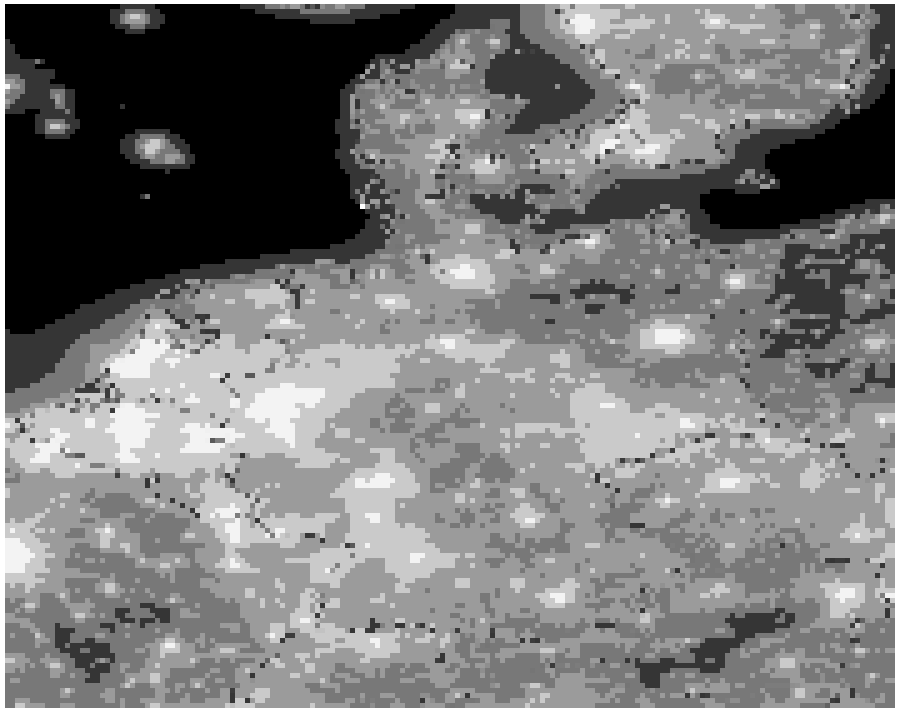} 
\caption[h]{Artificial sky brightness at sea level in Central Europe in V band  for aerosol content parameter $K\!=\!1$. }
\label{res6}
\end{figure}

Figures \ref{res1}-\ref{res6} show the maps of the  artificial sky brightness in Europe at sea level in V band.   The maps have been computed for clean atmosphere with an aerosol clarity $K=1$, corresponding at a vertical extinction of $\Delta m =0.33$ mag in V band, horizontal visibility $\Delta x=26$ km, optical depth $\tau=0.36$. 
Gray levels from black to white correspond to ratios between the artificial sky brightness and the natural sky brightness of: $<$11\%, 11\%-33\%, 33\%-100\%, 1-3, 3-9, $>$9. 
We limited our computations to zenith sky brightness even if our method allows determination of brightness in other directions. This would be useful to predict visibility in large territories of particular astronomical phenomena. A complete mapping of the artificial brightness of the sky of a site,  like Cinzano (1999a), but using satellite data instead of population data is possible (Cinzano 1999, in prep.). 

We are more interested in understanding and comparing light pollution distributions rather than predicting the effective sky brightness for observational purposes, so we  computed everywhere the artificial sky brightness at sea level, in order to avoid introduction of altitude effects in our maps.
We will  take account of altitudes  in a forecoming paper devoted to mapping the limiting magnitude and naked-eye star visibility which requires the computation of star-light extinction and natural sky brightness for the altitude of each land area.
We neglected the presence of mountains which might shield the light emitted from the sources from a fraction of the atmospheric particles along the line-of-sight of the observer.  
Given the vertical extent of the atmosphere in respect to the highness of the mountains, the shielding is not negligible only when the source is very near the mountain and both are quite far from the site (Garstang 1989, see also Cinzano 1999a,b). Earth curvature emphasizes this behaviour.

We calibrated the maps on the basis of both (i) accurate measurements of sky brightness together with extinction from the earth-surface and (ii) analysis of before-fly radiance calibration of  OLS-PMT.
Map calibration based on pre-fly irradiance calibration of OLS PMT require the knowledge, for each land area, of (a) the average vertical extinction $\Delta m$ during satellite observations and (b) the relation between the radiance in the choosen photometrical band and the radiance measured in the PMT spectral sensitivity range, which depends on the emission spectra.
The result of this calibration is  well inside the errorbar of the Earth-based calibration in spite of  the large uncertainties both in the extinction and in the average emission spectra. As soon as a large number of sky brightness measurements will be available, a better calibration will be possible.

We are extending this work to the rest of the World.

\acknowledgments

We are indebted to Roy Garstang of JILA-University of Colorado for his friendly kindness in reading and refereeing this paper, for his helpful suggestions and for interesting discussions.

%
%

\end{document}